\documentclass[11pt,conference]{IEEEtran}

 \usepackage{graphicx}
 \usepackage[cmex10]{amsmath}

\begin{document}
%
\title{Multi-scale Modularity in Complex Networks}

\author{\IEEEauthorblockN{Renaud Lambiotte}
\IEEEauthorblockA{Institute for Mathematical Sciences, Imperial College London\\
 53 Prince's Gate, London SW7 2PG, UK\\
Email: r.lambiotte@imperial.ac.uk}}

\maketitle

\begin{abstract}
We focus on the detection of communities in multi-scale networks, namely networks made of different levels of organization and in which modules exist at different scales. It is first shown that methods based on modularity are not appropriate to uncover modules in empirical networks, mainly because modularity optimization has an intrinsic bias towards partitions having a characteristic number of modules which might not be compatible with the modular organization of the system. We argue for the use of more flexible quality functions incorporating a resolution parameter that allows us to reveal the natural scales of the system. Different types of multi-resolution quality functions are described and unified by looking at the partitioning problem from a dynamical viewpoint. Finally, significant values of the resolution parameter are selected by using  complementary measures of robustness of the uncovered partitions. The methods are illustrated on a benchmark and an empirical network.
\end{abstract}


\IEEEpeerreviewmaketitle

\section{Introduction}

Many systems of current scientific interest are made of elements in interaction and can be represented as networks. Important examples include the Internet, telephone networks, collaboration networks, airline routes, but also a wide range of biological networks, such as food-webs, metabolic networks and protein interaction networks. The mathematical and empirical study of networks has emerged in the last decade as one of the fundamental building blocks in the wider study of complex systems \cite{review,evans,bocca}. One of the main reasons for this success is the possibility to analyze systems of a very different nature within a single framework. This approach allows to uncover similarities between the structures of various complex systems, which can reveal the existence of generic organization principles. A good example is the omnipresent multi-scale modular organization of complex networks, namely the fact that they are made of modules at different scales (see Fig.~1). Modules, also called communities, are defined as sub-networks that are locally dense even though the network as a whole is sparse \cite{GN}. The presence of modules at different scales is known to confer a crucial evolutionary advantage and to accelerate the emergence of complex systems by providing stable intermediate building blocks \cite{simon}. 

\begin{figure}[t]
\includegraphics[width=0.47\textwidth]{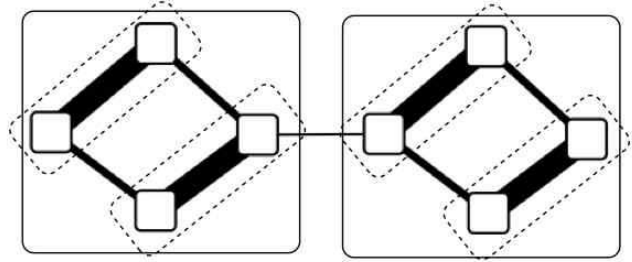}

\caption{Sketch of a multi-scale weighted network. The width of the links is proportional to their weight. This network is clearly made of modules at different scales: $8$ single nodes, $4$ pairs of strongly connected nodes, $2$ groups of $4$ nodes and the system as whole. This multi-scale network is hierarchical as modules at one level are nested into modules at the next level, but this is not necessarily the case, i.e., some multi-scale networks are not hierarchical.}\label{fig0}
\end{figure}

The capacity to collect large data-sets of relational data has radically changed the way networks are considered and has led to the development of statistical methods for the description of their multi-scale topology and the detection of significant connectivity patterns. A powerful set of methods
consists in uncovering the modules present in the network \cite{fort,revporter}. This identification has the advantage of providing a coarse-grained representation of the system, thereby allowing to sketch its organization and to identify sets of nodes that are likely to have hidden functions or properties in common. Most community detection methods find a partition of the nodes into communities, where most of the links are concentrated within the communities.  Each node is assigned to one and only one community, i.e., partitions are not compatible with overlapping communities \cite{overlap1,overlap2}. At the heart of most partitioning methods, there is a mathematical definition for what is thought to be a good partition. Once this quality function has been defined, different types of heuristics can be used in order to find, approximatively, its optimal partition, i.e., to find the partition having the highest value of the quality function.

 In this article, we first describe different multi-resolution quality functions, namely quantities incorporating a resolution parameter allowing to tune the characteristic size of modules in the optimal partition. We show that these quantities are linearised versions of a quality function called stability \cite{JC}, which is based on the exploration of the network by a random walker at different time scales. Finally, we focus on the optimization of these quality functions and on the important issue of detecting significant values of the resolution parameter in practical applications. The methods are successfully tested on a benchmark and on a real-world network.

\section{Modularity and its limitations}

Let $A$ be the adjacency matrix of a weighted, undirected network. $A$ is therefore symmetric and $A_{ij}$ is the weight of the link between $i$ and $j$. The strength of node $i$ is defined as $k_i \equiv \sum_{j} A_{ij}$; $m \equiv \sum_{i,j} A_{ij}/2$ is the total weight in the network. If the network is unweighted, $k_i$ and $m$ are the degree of node $i$ and the total number of links respectively. The quality of the partition of a network is a function of the adjacency matrix $A$ and of the partition $\mathcal P$ of the nodes into communities. The widely-used modularity \cite{NG} of a partition $\mathcal P$ measures if links are more abundant within communities than would be expected on the basis of chance
\begin{align}
Q &= \mbox{(fraction of links within communities)} \nonumber\\
  &\qquad{} - \mbox{(expected fraction of such links)}
\label{eq:modDeff}
\end{align}
and reads  
\begin{equation}
\label{modularity}
Q = {1\over2m} \sum_{C \in \mathcal{P}} \sum_{i,j \in C} \biggl[ A_{ij} - P_{ij} \biggr],
\end{equation}
where  $i,j \in C$ is a summation over pairs of nodes $i$ and $j$ belonging to the same community $C$ of $\mathcal P$ and therefore counts intra-community links.
The null hypothesis is an extra ingredient in the definition and is incorporated in the matrix $P_{ij}$. $P_{ij}$ is the expected weight of a link between nodes $i$ and $j$  over an ensemble of random networks with certain constraints. These constraints correspond to known information about the network organization, i.e., its total number of links and nodes, which has to be taken into account when assessing the relevance of an observed topological feature. Two standard choices for the corresponding null models are
\begin{eqnarray}
P_{ij}=\langle k\rangle^2/2m, & \text{then $Q \equiv   Q_{\rm unif}$} 
\end{eqnarray}
where $\langle k\rangle=2m/N$ is the average strength and the only constraint is thus the total weight in the network, and
\begin{eqnarray}
P_{ij}= k_i k_j/2m, &  \text{ then $Q \equiv  Q_{\rm conf}$.}
\end{eqnarray}
where randomized networks now preserve the strength of each node. The latter null model is usually preferred because it takes into account the degree heterogeneity of the network~\cite{lastn}. More complicated null models can in principle be constructed in order to preserve other properties of the network under consideration \cite{gen1,gen2}.

It is interesting to note that $Q_{\rm unif}$ and $Q_{\rm conf}$ are naturally related to the combinatorial Laplacian $L^{(C)}_{ij}=A_{ij} - k_i \delta_{ij}$ and the (normalized) Laplacian $L_{ij}=A_{ij}/k_j - \delta_{ij}$  respectively\footnote{ Strictly speaking, the normalized Laplacian of a network is $L^{'}_{ij}=A_{ij}/(k_i^{1/2} k_j^{1/2}) - \delta_{ij}$, but $L$  and $L^{'}$ are equivalent by similarity as $L^{'}_{ij} = k_i^{-1/2} L_{ij} k_j^{1/2}$.}, and, more generally, to the dynamics induced by these operators (see section \ref{stability}). For $Q_{\rm conf}$, this relation is particularly clear after expressing modularity in terms of the (right) eigenvectors $v_{\alpha}$ of $L_{ij}$, i.e., $v_{\alpha}$ satisfy  $\sum_j L_{ij} v_{\alpha,j}=\lambda_\alpha v_{\alpha,i}$. Without loss of generality, we assume that $\lambda_{1} > \lambda_{2} \geq ... \geq \lambda_{\alpha} \geq ... \geq \lambda_{N}$. The dominant eigenvector $v_1$ of eigenvalue $\lambda_1=0$ is given by $v_{1;i}=k_i/2m$ and is unique if the network is connected. By using a spectral decomposition of $L_{ij}$, one finds \cite{JC}
\begin{eqnarray}
\label{lll}
Q_{\rm conf} =  \sum_{\alpha=2}^{N} \frac{\lambda_{\alpha}+1}{2m} \sum_C \sum_{i,j \in C} v_{\alpha;i} v_{\alpha;j},
\end{eqnarray}
where the contribution of the dominant eigenvector $v_1$ and the null model have cancelled each other out.

The optimization of modularity has the advantage of being performed without a priori specifying  the number of modules nor their size. This procedure has been shown to produce useful and relevant partitions in a number of systems \cite{newman_modul_PNAS}. Unfortunately, it has also been shown that modularity suffers from several limitations, partly because modularity optimization produces one single partition, which is not satisfactory when dealing with multi-scale systems. Related to this issue, there is the so-called resolution limit of modularity \cite{FB}, namely the fact that modularity is blind to modules smaller than a certain scale. This point originates from the bias of modularity towards modules having a certain scale which might not be compatible with the system architecture. This incompatibility also makes modularity inefficient in practical contexts as it may lead to a high degeneracy of its landscape \cite{Good}, i.e., the existence of several distinct partitions having a modularity close to the optimum, which implies that approximate solutions of the optimization problem are very dissimilar and that a partition derived from modularity optimization has to be considered with caution.

\section{Multi-scale methods}

\subsection{Local maxima of modularity}
Different methods have been proposed to go beyond modularity optimization. A first set of methods looks for local maxima of the modularity landscape in order to uncover 
partitions at different scales \cite{sales}. A good example is the so-called ÒLouvain methodÓ, which is a greedy method taking advantage of the hierarchical organization of complex networks in order to facilitate the optimization of modularity \cite{Blondel}. This heuristic performs the optimization in a multi-scale way: by comparing the communities first of adjacent nodes, then of adjacent groups of nodes found in the first round, etc. It has been shown in several examples that modularity estimated by this method is close to the optimal value obtained with slower methods, but also that intermediate partitions are meaningful and correspond to communities at intermediate resolutions \cite{meunier,livre}. This approach has the advantage of being fast, but it lacks theoretical foundations and is not able to uncover coarser partitions than those obtained by modularity optimization. Moreover, it may produce hierarchies even when the system is single-scale or, worse, completely random (see \cite{meunier} for a discussion of how to deal with this issue). 

\subsection{Multi-resolution quality functions}
Another class of methods is based on multi-scale quality functions. These quality functions incorporate a resolution parameter allowing to tune the characteristic size of the modules in the optimal partition and aim at uncovering modules at the true scale of organization of a network, i.e., not at a scale imposed by modularity optimization. The two most popular multi-scale quality functions are ad-hoc, parametric generalizations of modularity. A first quantity is the parametric modularity introduced by Reichardt and Bornholdt \cite{reichardt0,reichardt}
\begin{equation}
\label{reichardt}
Q_\gamma = {1\over2m} \sum_{C \in \mathcal{P}} \sum_{i,j \in C} \biggl[ A_{ij} - \gamma P_{ij} \biggr],
\end{equation}
which is usually defined for the configuration null model $P_{ij}= k_i k_j/2m$ and mainly consists in changing the effective size of the system $m_{\rm eff}=m/\gamma$. The optimization of $Q_\gamma$ leads to larger and larger communities in the optimal partition when $\gamma$ is decreased. This approach makes use of the  size dependence of modularity: because of the factor $1/2m$ in the null model, modularity depends on the total size of the network and not only on its local properties\footnote{ In a nutshell, this size dependence originates from a choice of null model where each pair of nodes $i$ and $j$ can be connected, given a certain number of available links $m$ in the system, whatever the distance between $i$ and $j$ in the network. Local null models where pairs of nodes are randomly connected only within a finite radius of interaction are expected to suppress this effect.}. Decreasing $m_{\rm eff}$ (increasing $\gamma$) increases the expected weight $\gamma P_{ij}$ of a link between $i$ and $j$, which makes it less advantageous to assign $i$ and $j$ to the same community (because $A_{ij} - \gamma P_{ij}$ decreases).  

An alternative approach proposed by Arenas et al. \cite{stability1} keeps modularity unchanged but modifies the network by adding self-loops to the original network. This approach therefore consists in optimizing 
\begin{equation}
\label{arenas}
Q_r = Q(A_{ij} + r I_{ij}).
\end{equation}
As expected, increasing $r$ has a tendency to decrease the size of the communities and the optimal partition of $Q_{\infty}$ is made of single nodes. Even if increasing $\gamma$ and $r$ has, qualitatively, the same effect on the characteristic size of the communities, one should keep in mind that $Q_\gamma$ and $Q_r$ are in general optimized by different partitions, except if the network is regular and the resolution parameters verify $\gamma=1+r/\langle k \rangle$. It is also interesting to note that the quality function (\ref{arenas}) was first proposed in order to preserve the eigenvectors of the adjacency matrix, as the eigenvectors of $A_{ij} + r I_{ij}$ and $A_{ij}$ are obviously the same. From a partitioning viewpoint, however, 
the eigenvectors of $A_{ij}$ do not matter as much as the eigenvectors of the combinatorial Laplacian $L_{ij}^{(C)}$ \cite{fiedler} and the 
normalized Laplacian $L_{ij}$ \cite{shimalik}. Moreover, modularity is related to the eigenvectors of the Laplacian and not of the adjacency matrix, see (\ref{lll}). These observation suggest to adapt the unfitting quality function (\ref{arenas}) and to optimize the modularity of a modified adjacency matrix preserving the eigenvectors of $L_{ij}$. This can readily be done by adding strength-dependent self-loops to the nodes 
\begin{equation}
A_{ij}^{'} =A_{ij} + r \frac{k_i}{\langle k \rangle} \delta_{ij},
\end{equation}
and by optimizing the quality function
\begin{equation}
\label{general}
Q^{'}_r \equiv Q(A_{ij} + r \frac{k_i}{\langle k \rangle} \delta_{ij}).
\end{equation}
This quality function is equivalent, up to a linear transformation, to $Q_\gamma$ for any network, i.e., not only for regular networks, with $\gamma=1+r/\langle k \rangle$, thereby providing two alternative interpretations to resolutions parameters. 

\subsection{Stability}

\label{stability}
The multi-resolution quality functions defined in the previous section have been successfully tested on multi-scale benchmark and empirical networks \cite{reichardt,ronhovde,mason}. They have the further advantage of being mathematically very similar to modularity and of being optimized by modularity optimization algorithms with minimum  code development. Unfortunately, the introduction of a resolution parameter, $\gamma$ or $r$, feels like a trick and lacks theoretical ground. In order to define a resolution parameter in a more satisfying way and, as we will see, to provide a more solid foundation to $Q_\gamma$ and $Q^{'}_r$,  we look at communities from a different angle, not from a combinatorial point of view, where intra-community links are counted as in (\ref{modularity}), but from a dynamical point of view. 

Our starting point is the following: a flow taking place on a network is expected to be trapped for long times in good communities before being able to escape \cite{rosvall,JC}. This argument suggests to measure the quality of a partition in terms of the persistence of flows taking place on the network \cite{JC,lambi}. Without loss of generality, we describe a stationary Markov process $\mathcal M$ as a random walk process. Under the condition that $\mathcal M$ is ergodic, i.e., any initial configuration asymptotically reaches the unique stationary solution, stability is defined as
\begin{align}
R_{\mathcal M}(t) &= \mbox{(probability for a random walker to be in the } \nonumber\\
& \mbox{ same community initially and at time $t$)} \nonumber\\
  & - \mbox{(probability for two independent random }  \nonumber\\
  & \mbox{walkers to be in the same community)}
\label{eq:stabDeff}
\end{align}
when the system is at equilibrium.

In order to clarify this general concept, let us focus on a generic Markov process \cite{lambi}, namely a continuous-time random walk where waiting times are independent, identical Poisson processes. The density of random walkers on node $i$ at time $t$, denoted by $p_{i}(t)$, evolves according to the rate equation
\begin{equation}
\label{ctrw}
\dot{p}_{i} = \sum_{j} \frac{A_{ij}}{k_j} p_{j} - p_i,
\end{equation}
where $\frac{A_{ij}}{k_j} - \delta_{ij} \equiv L_{ij}$ is the Laplacian operator described above. In this unbiased process, a walker located at $j$ follows a link going to $i$ with a probability proportional to $A_{ij}$. If the network is connected, the stationary solution is unique and given by the dominant eigenvector of $L$, namely $p_i^*=k_i/2m$.
By definition (\ref{eq:stabDeff}), the stability of a partition associated to the Markov process (\ref{ctrw}) is 
\begin{equation}
\label{Rt}
R_{\rm NL}(t) =   \sum_C \sum_{i,j \in C} \biggl[ \left( e^{t L} \right)_{ij} {k_j\over2m} - {k_i\over2m} {k_j\over2m} \biggr],
\end{equation}
where ${\rm NL}$ stands for Normalized Laplacian. This expression clearly shows that stability depends on time. The quality of a partition is thus measured differently at different time scales and  is, in general, optimized by different partitions when time is tuned, thereby leading to a sequence of optimal partitions. 

By looking at limiting values of $t$, one can show that time acts as a resolution parameter \cite{JC,lambi}. 
As time grows, the characteristic size of the communities is thus adjusted to reveal the possible multi-scale organization of the system. 
In the limit $t \rightarrow 0$, keeping linear terms in $t$ in the expansion of $R_{\rm NL}(t)$ leads to
\begin{equation}
\label{Rsmall}
R_{\rm NL}(t) \approx  (1-t)\, R_{\rm NL}(0) + t \, Q_{\rm conf} \equiv Q_{\rm NL}(t),
\end{equation}
which is equivalent up to a linear transformation to $Q_\gamma$ and $Q^{'}_r$ when $P_{ij}=k_i k_j/2m$ (with $t=1/\gamma$, $t=\langle k \rangle/(r+\langle k \rangle)$). These multi-resolution quality functions can therefore be seen as a simple linear approximation of $R_{\rm NL}(t)$, which provides a physical interpretation to the resolution parameter $r$ and $\gamma$, i.e., the inverse of the time used to explore the network. It is also interesting to note that the configuration null model naturally emerges from the definition of stability and from the dynamics (\ref{ctrw}). Interestingly, other null models, including the uniform null model, are associated to other random walk processes \cite{lambi}.
 In the limit $t \rightarrow \infty$, making use of the spectral decomposition of $L$, stability simplifies as
\begin{equation}
\label{longtime}
R_{\rm NL}(t) \approx   \frac{1}{2m} e^{t \lambda_{2}} \sum_C \sum_{i,j \in C}  v_{2;i} v_{2;j},
\end{equation}
where it is assumed that the second dominant eigenvalue $\lambda_{2}$ of $L$ is not degenerate and $v_2$ is its corresponding (right) eigenvector. $R_{\rm NL}(t)$ is therefore maximized by a partition into two communities in accordance with the normalized Fiedler eigenvector~\cite{shimalik}.

$R_{\rm NL}(t)$ differs from modularity in several ways \cite{lambi}. However, one can show that $R_{\rm NL}(t)$ is always equal to the modularity $Q_{\rm conf}$ of a time-dependent weighted network whose adjacency matrix is $X_{ij}(t)=\left(e^{t L} \right)_{ij} k_j$:
\begin{equation}
\label{relation}
R_{\rm NL}(t) \equiv Q_{\rm conf}(X_{ij}(t)).
\end{equation}
 This new network is symmetric if the original network is symmetric and the weight on its links corresponds to the number of walkers going from $j$ and $i$ in time $t$, when the system is at equilibrium. By construction, $X(t)$ is more and more extended when $t$ is increased. The optimization of its modularity is therefore expected to uncover larger communities. After noting that the Laplacian of $X_{ij}$ and the Laplacian of $A_{ij}$ have the same eigenvectors, (\ref{relation}) emphasizes that $R_{\rm NL}(t)$ naturally fits the arguments used to define $Q^{'}_r$, see (\ref{general}).

\section{Optimization, robustness and selection of significant scales}

Let us now discuss the practical side of this work, namely the detection of multi-scale communities in large empirical networks. In what follows, we will focus on the optimization of $Q_{\rm NL}(t)$ (and equivalently of $Q_\gamma$ and $Q^{'}_r$), while keeping in mind that the optimization of the full stability $R_{\rm NL}(t)$ can be performed by using spectral or greedy methods \cite{lambi}. Depending on the size of the network under consideration, generalizations of different modularity optimization heuristics can be used in order to optimize $Q_{\rm NL}(t)$, such as Simulated Annealing, Spectral Methods or Greedy Methods for small ($N<10^2$), intermediate ($N \sim 10^3$) and large  ($N> 10^4$) sparse networks respectively. In the following, we perform the optimization of $Q_{\rm NL}(t)$ by using a generalization of the Louvain method \cite{Blondel} mentioned above\footnote{ Codes are available on {\it http://www.lambiotte.be.}}. One should stress that the outcome of the algorithm is deterministic, except in the initial ordering (labeling) of the nodes. This implies that different optimal partitions (local maxima of $Q_{\rm NL}(t)$) can in principle be uncovered when the initial ordering is changed.

Partitions at different values of $t$ are found independently by optimizing $Q_{\rm NL}(t)$, thereby producing a sequence of partitions that are optimal at different scales.
However, one expects that only a small number of these partitions are significant, which raises another question: how can one select the most significant partitions, or equivalently the most significant scales of description of the network? It is ironical to note that we are thus confronted with a problem similar to the one that initially led to the definition of modularity. Modularity was indeed first proposed to find the best partition in a nested hierarchy of possible community divisions  \cite{NG}. As we have argued before and will show below on an example, modularity does not appropriately detect important scales of description. 

In order to address this problem, it has recently been proposed to look for robust partitions, where robustness has been defined differently by different authors. This approach formalizes the intuitive idea that a significant partition should not be altered by small modifications. A standard measure to compare two partitions $\mathcal{P}_1$ and $\mathcal{P}_2$ is the so-called normalized variation of information $\hat{V}(\mathcal{P}_1,\mathcal{P}_2)$  \cite{meila}, which is a number between 0 and 1 and is equal to 0 only when the partitions are identical. Three types of modifications have been proposed:

{\bf Modifying the network} by reshuffling a fraction of the links \cite{karrer} or randomly perturbing the weight of the links \cite{ros2}. In the following, we implement the second approach by randomly adding $\pm 10 \%$ to the weight of the links. In practice, we optimize $Q_{\rm NL}(t)$ for $K$ different realizations of the perturbed network for each value of $t$, by always using the same node ordering. The robustness of partitions at time $t$ is given by 
\begin{equation}
\langle V \rangle_{\rm net}(t)=\frac{2}{K (K-1)}\sum_{k=1}^{K}\sum_{k^{'}=k+1}^{K}\hat{V}(\mathcal{P}_k(t),\mathcal{P}_{k^{'}}(t)),
\end{equation}
where $\mathcal{P}_k(t)$ is the optimal partition of the $k^{\rm th}$ realization of the perturbed network at time $t$. In this approach, a scale is significant when a small modification of the network does not alter too much the partition found by the optimization algorithm.

{\bf Modifying the optimization algorithm} by taking advantage of the dependence of the algorithm on its initial condition, i.e., the node ordering \cite{ronhovde,meunier}. To do so, we optimize $Q_{\rm NL}(t)$ of the original network $T$ times by attributing a different, random ordering to the nodes. Robustness at time $t$ is 
\begin{equation}
\langle V \rangle_{\rm algo}(t)=\frac{2}{T (T-1)}\sum_{i=1}^{T}\sum_{i^{'}=i+1}^{T}\hat{V}(\mathcal{P}_i(t),\mathcal{P}_{i^{'}}(t)),
\end{equation}
where $\mathcal{P}_i(t)$ is now the optimal partition when using the $i^{\rm th}$ random ordering of the nodes at time $t$. In this approach, robustness measures the size of the basin of attraction\footnote{ The deterministic optimization process can be seen as a trajectory in the space of partitions toward a (ideally global) maximal partition. The optimization always starts from the finest partition, where each node belongs to its own community, but its next steps depend on the ordering on the nodes. The basin of attraction of an uncovered partition is the set of initial node orderings that lead to it.} of the optimal partitions.

{\bf Modifying the quality function} by tuning the resolution parameter \cite{mason,lambi}. To do so, we perform one optimization of $Q_{\rm NL}(t)$ for each $t$, while keeping the node ordering fixed throughout the different values of $t$. The robustness of partitions at time $t$ is 
\begin{equation}
\langle V \rangle_{\rm QF}(t)=\frac{1}{\Delta}\sum_{\tau=1}^{\Delta} \hat{V}(\mathcal{P}(t),\mathcal{P}(t+\tau)),
\end{equation}
where $\mathcal{P}(t)$ is the optimal partition at time $t$. In this approach, robustness corresponds to the persistence of an optimal partition over long periods of time, and to the fact that optimal partitions are weakly altered by tuning $t$.

In each case, robustness is related to the ruggedness of the quality function landscape. Lack of robustness corresponds to high degeneracy, namely to the existence of incompatible partitions that are local maxima of $R_{NL}(t)$ with a value close to the global maximum. Significant partitions are uncovered by identifying values of the resolution parameter where these measures of robustness are significantly low.

\section{Tests of the method}

\begin{figure}[t]
\includegraphics[width=0.45\textwidth]{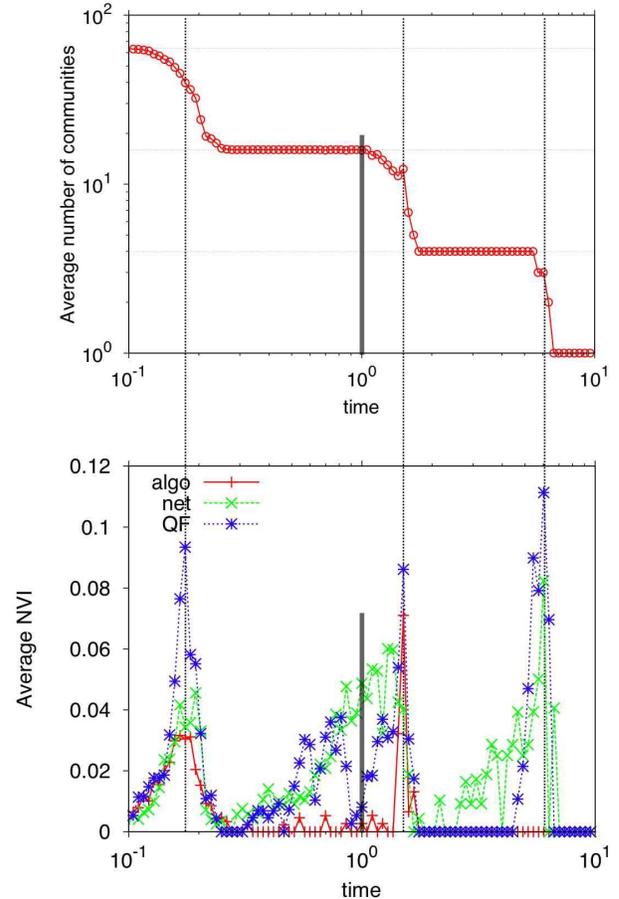}

\caption{Analysis of a hierarchical network with $N=640$ nodes and $3$ known hierarchical levels (natural partitions into $4$, $16$ and $64$ modules). In the upper figure, we plot the average number of modules as a function of the resolution parameter $t$ when performing $100$ optimizations with random orderings (associated to the definition of $\langle V \rangle_{\rm algo}(t)$, see main text). In dashed line, we plot the expected numbers of modules in the natural partitions. In the lower figure, we plot the measures of robustness $\langle V \rangle_{\rm algo}(t)$, $\langle V \rangle_{\rm net}(t)$ and $\langle V \rangle_{\rm QF}(t)$. Natural partitions are robust and associated to low values of $\langle V \rangle$, while in-between values of $t$ are characterized by peaks of $\langle V \rangle$ (except for $\langle V \rangle_{\rm algo}(t)$ at the jump from $4$ modules to $1$ module). Vertical lines indicate peaks in $\langle V \rangle$ and the behavior of the system at $t=1$ (modularity).}\label{fig1}
\end{figure}

\begin{figure}[t]
\includegraphics[width=0.45\textwidth]{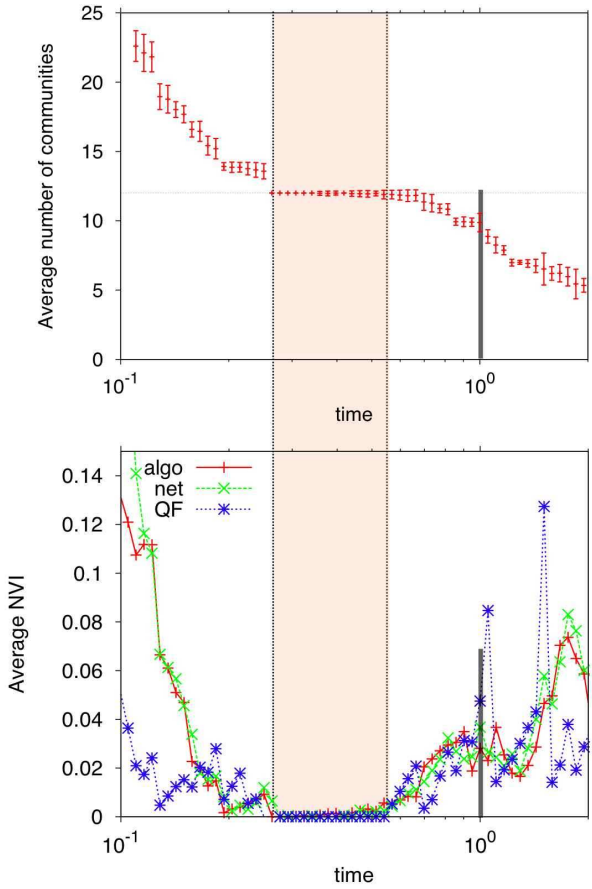}

\caption{Analysis of the college football network.  In the upper figure, we plot the average number $N_t$ of modules and its standard deviation $\sigma_t$ when performing $100$ optimizations with random orderings. In dashed line, we plot the number of modules in the expected partition ($12$ communities). In the lower figure, we plot the measures of robustness $\langle V \rangle_{\rm algo}(t)$, $\langle V \rangle_{\rm net}(t)$ and $\langle V \rangle_{\rm QF}(t)$. The only time window where $\langle V \rangle$ vanishes corresponds to a partition into $12$ communities. Interestingly, $t=1$ (indicated by a vertical line) is not particularly robust, i.e., modularity does not uncover modules at the appropriate scale. As shown in the upper figure, for each value of $t$, the numbers of modules in the $100$ optimal partitions are close to their average $N_t$ (small values of $\sigma_t$) even when the partitions are not robust.}\label{fig2}
\end{figure}

In this section we test these ideas by focusing on  a computer-generated network and a real-world network for which the community structure is already known. In each case we find that the method reliably detects the known community structure and reveals the important scales of description. 

\subsection{Hierarchical benchmark network} This randomly-generated network is made of $640$ nodes with $3$ known hierarchical levels: small modules of $10$ nodes nested in medium-size modules of $40$ nodes themselves nested in large modules of $160$ nodes \cite{sales}. The expected number of links across modules and therefore the sharpness of the modules is tuned by a single parameter $\rho$, $\rho=1.0$ in this example. We focus on one single realization of this random network. In order to evaluate $\langle V \rangle_{\rm net}(t)$, $\langle V \rangle_{\rm algo}(t)$ and $\langle V \rangle_{\rm QF}(t)$, we use $K=10$, $T=10$ and $\Delta=5$. As shown in Fig.~2, the method clearly uncovers the correct scales of description and only those scales. The natural partitions into $4$, $16$ and $64$ modules respectively are characterized by regions of $t$ where $\langle V \rangle(t) \ll 1$. Moreover, these regimes are clearly separated by peaks of $\langle V \rangle(t)$, i.e., values of the resolution parameter where the algorithm finds conflicting partitions. It is interesting to note that $\langle V \rangle_{\rm net}(t)$, $\langle V \rangle_{\rm algo}(t)$ and $\langle V \rangle_{\rm QF}(t)$ have similar but non-identical patterns, which suggests to combine the use of different notions of robustness in order to improve the detection of significant partitions. Partitions uncovered by modularity ($t=1$) optimization have, on average, $15.9 \pm 0.3$ modules when measured over $100$ optimizations with random orderings.

\subsection{College football}  This real-world network is made of $115$ football teams that are connected if they have played a regular-season game \cite{GN}. Because games are more frequent between members of the same conference than between members of different conferences, one expects a natural partition into $12$ communities, corresponding to the $12$ conferences of the championship. For this network, we use the parameters $K=100$, $T=100$ and $\Delta=5$. One observes (see Fig.~3) a clear plateau where partitions are made of $12$ communities and where the three versions of robustness are vanishingly small. It is interesting to note that $t=1$ does not belong to this plateau. Modularity optimization thus fails to uncover a robust partition and provides an inappropriate representation of the system. This intrinsic problem of modularity has already been observed in benchmarks \cite{sasa} and empirical networks \cite{JC}.

\section{Discussions}

In this article, we have focused on the detection of non-overlapping modules in multi-scale networks. These networks are made of different levels of organization and are typically (but not necessarily) hierarchical, in the sense that the system is made of modules, which themselves are made of sub-modules, etc.
We have shown that modularity optimization is not a satisfactory method to uncover modules in general, because modularity optimization reveals communities at scales that are not automatically compatible with the system organization. It is therefore necessary to incorporate a resolution parameter to modularity in order to adjust the characteristic size of the modules and to uncover the true modular organization of a network. Three different multi-resolution quality functions $Q_\gamma$, $Q^{'}_r$ and $Q_t$ have been presented. They are all equivalent up to a linear transformation and include modularity as a particular case when the resolution parameters are $\gamma=1$, $r=0$ and $t=1$. It is important to keep in mind that multi-resolution quality functions have the same limitations as modularity when the resolution parameter is fixed \cite{jari} and that  the possibility to tune this parameter is essential. No value of the resolution parameter is a priori better than another one and additional tests are therefore needed to uncover significant scales of description. Contrary to what is usually believed, modularity is thus an ordinary instance in the set of multi-resolution quality functions and there is no deductive reason to prefer it. Important values of the resolution parameter have instead to be selected by considering the robustness of the detected partitions. Our analysis suggests to develop proper statistical tests and to combine the information obtained from different measures of robustness in order to better comprehend the modular organization of complex networks.

\section*{Acknowledgment}

I would like to thank M.\ Barahona, J.-C.\ Delvenne, T.\ Evans, S. Fortunato and D. Meunier for fruitful discussions, E. Landuyt for proof-reading and J.-L. Guillaume for providing the C++ code of the Louvain method for modularity optimization\footnote{ http://sites.google.com/site/findcommunities/} \cite{Blondel}. This work has been supported by the UK EPSRC and was conducted within the framework of COST Action MP0801 Physics of Competition and Conflicts.



%

\end{document}